\documentclass[a4paper]{jpconf}
\usepackage{graphicx}
\begin{document}
\title{Alpha particle condensation and nuclear rainbow scattering }

\author{S Ohkubo$^1$ and Y Hirabayashi$^2$}

\address{ $^1$Department of Applied Science and Environment, Kochi Women's University, 
Kochi 780-8515, Japan
}

\address{$^2$Information Initiative Center, Hokkaido University, Sapporo 060-0811, Japan
}

\ead{shigeo@cc.kochi-wu.ac.jp}

\begin{abstract}
It is shown that  a dilute property of an $\alpha$ particle condensate
 can be seen in the Airy structure of nuclear rainbow and prerainbow scattering.
 The dilute property of the Hoyle state of $^{12}$C with a developed
 $\alpha$ cluster structure is  discussed by studying  refractive 
$^3$He+$^{12}$C and $\alpha$+$^{12}$C scattering.
\end{abstract}

\section{Introduction}
   Bose-Einstein condensation of   $\alpha$ particles  is a challenging subject
and recently Tohsaki {\it et al.} \cite{Tohsaki2001} conjectured that 
the  $0_2^+$(7.65 MeV) state of $^{12}$C, the Hoyle state, has a dilute density 
distribution due to Bose-Einstein condensation 
of three $\alpha$ particles.
The large radius of the Hoyle state due to Bose-Einstein condensation has
 not been measured directly. 
 When absorption is incomplete a projectile which  penetrates deep into
 the nucleus can carry the information about the  
nuclear potential. 
When the nuclear radius 
is very large due to Bose-Einstein condensation,  refractive scattering would greatly differ from the scattering
 from a nucleus with a normal size.
Our approach in this paper is to use a refractive property of the nucleus as 
a lens.

The sensitivity of  the rainbow angle and its associated   Airy structure 
on refractive index $n$ is seen   in the meteorological  rainbow. In Fig.~1
 Mie scattering from a small water droplet is shown. For a red wave ($n=$1.33)  
the primary rainbow appears  at about  137.7$^\circ$ and for a violet wave ($n$=1.34)
 at 139.5$^\circ$. As the refractive index  increases the rainbow angle  and  
  the  associated Airy structure is shifted to backward. 

On the other hand,  different from a  meteorological   rainbow, a nuclear rainbow 
is Newton's zero-order
 rainbow \cite{Michel2002} and has no internal reflection in the medium.
However the sensitivity 
of the rainbow angle and its associated  Airy structure on the refractive index, i.e.
 nuclear potential,  can be  seen as well.
In fact, nuclear rainbow  scattering has been  powerful in the study of nucleus-nucleus
 interaction potential  when
 absorption is incomplete \cite{Khoa2007}. 
Nuclear rainbow scattering typically occurs at the  high energy region where a 
semiclassical
 picture holds.
In optics  refractive index is related to the optical potential  as follows:
 \begin{equation}
 n(r) = \sqrt{1-\frac{V(r)}{E_{c.m.}}}. 
\end{equation} 
This shows that the refractive index $n(r)$ becomes large as the incident energy
 becomes lower.
 Rainbow scattering has mostly been  
 studied for elastic scattering in optics and also in nuclear physics. Inelastic 
rainbow scattering has not been very clear theoretically and experimentally.
Recently it
 has been  shown that  the evolution of the Airy structure in 
inelastic scattering is  similar to elastic scattering \cite{Michel2004B}.

 The concept of a  prerainbow  has been proposed to understand the 
refractive phenomenon at the  lower energy region where rainbow-like "Airy" 
oscillations  without a falloff of the cross sections
 appears  in the angular distributions \cite{Michel2002}.
This prerainbow shows  clear  oscillations, which are understood as the 
interference between the subamplitude of the farside component of the internal waves,
which penetrate the barrier deep into the inner part of the potential 
and are  reflected at the most internal turning point, and the farside component of the 
barrier waves, which are reflected at the  barrier of the potential at the surface.
 We show that strong refraction of the Hoyle state can be seen  in the prerainbow
 oscillations in inelastic $^3$He+$^{12}$C scattering and inelastic 
rainbow scattering of an   $\alpha$ particle from $^{12}$C \cite{Ohkubo2007,Ohkubo2004}. 

\begin{figure}[h]
\begin{center}
%\begin{minipage}[b]{18pc}
\includegraphics[width=17pc]{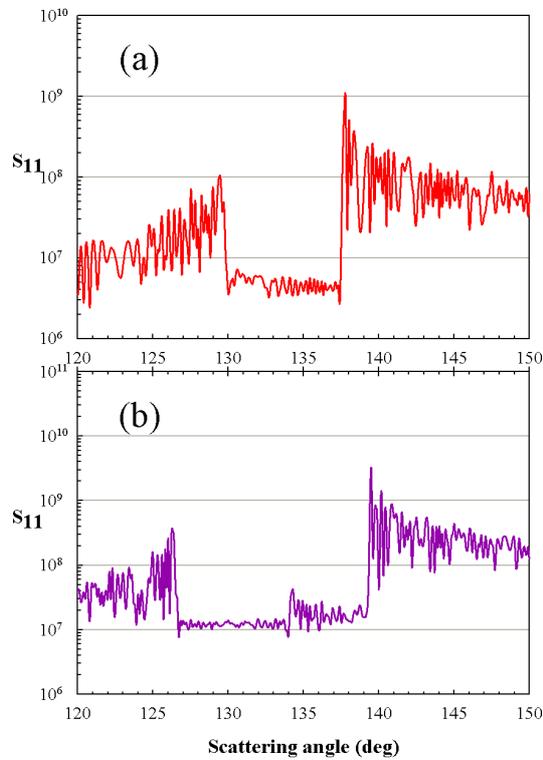}
%\end{minipage}\hspace{1.5pc}
%\begin{minipage}[b]{18pc}
\caption{Mie scattering intensity profiles for  a  water droplet (radius 3 mm) 
 at two visible wavelengths. (a)   
 $\lambda$=0.65 $\mu$m (red, $n=1.33$) and (b)
 $\lambda$=0.40 $\mu$m (violet,  $n$=1.34). 
 }
%\end{minipage}
\end{center}
\label{fig 1Mie scattering}
\end{figure}

 \section{Double folding model}\label{model:sec2}
Kamimura {\it et al.} \cite{Kamimura1981} studied the structure of $^{12}$C 
 in the Resonating Group Method
by using a three $\alpha$ particle cluster model. The obtained cluster wave functions
 reproduce many experimental 
data  such as energy levels, $\alpha$ decay widths, electric transition 
probabilities, electron  scattering form factors and have been used successfully in 
many reactions involving  $^{12}$C.    The wave function for the Hoyle state
 is almost completely equivalent  to
the  Bose-Einstein condensate wave function \cite{Funaki2003}.
Therefore in the following coupled channel analysis with the double folding model
 we use the wave functions  of Kamimura {\it et al.}  \cite{Kamimura1981}.
The double folding potential is given  by 
\begin{equation}
V_{ij}({\bf R}) =
\int \rho_{00} ({\bf r}_{1})\;
     \rho_{ij}^{\rm (^{12}C)} ({\bf r}_{2})\;
v_{\rm NN} (E,\rho,{\bf r}_{1} + {\bf R} - {\bf r}_{2})\;
{\rm d}{\bf r}_{1} {\rm d}{\bf r}_{2}  \; ,
\end{equation}
\noindent
where $\rho_{00} ({\bf r})$ is the ground 
state density
of  a projectile, while $v_{\rm NN}$ denotes
 the density-dependent M3Y effective interaction (DDM3Y) \cite{Kobos1984}.
$\rho_{ij}^{\rm (^{12}C)}  ({\bf r})$ represents the diagonal 
($i=j$) or transition ($i\neq j$) nucleon density of $^{12}$C
calculated by Kamimura {\it et al.} \cite{Kamimura1981}.
The density distribution of  $^3$He is taken from  Cook {\it et al} \cite{Cook1981}.
 In the analysis  we  introduce the normalization factor 
 $N_R$ for 
 the real part of the potential and phenomenological
 imaginary potentials with a  Wood-Saxon form factor  
and a derivative of the  Wood-Saxon form factor   for each channel.

 \section{Analysis of $^3$He+$^{12}$C scattering}\label{analysis:sec3}
\par

We study  elastic and inelastic $^3$He+$^{12}$C scattering
  in the microscopic coupled channel method by taking into 
 account simultaneously the  0$^+_1$ (0.0 MeV), $2^+$ (4.44 MeV),
 0$^+_2$ (7.65 MeV),  and 3$^-$ (9.63 MeV) states of $^{12}$C.
 Calculated angular distributions   
 at 34.7 MeV and    72 MeV are shown in Fig.~2  in comparison 
    with the experimental data \cite{Fujisawa1973,Demyanova1992}.
The normalization factor $N_R=$1.28 is used and the imaginary potential parameters are
adjusted to fit the data.
 The calculation reproduces the  experimental  angular
 distributions for the ground state and the Hoyle state
 as well as the $2^+$ 
  and 3$^-$ states  \cite{Ohkubo2007}.
\begin{figure}[tbh]
\begin{center}
\includegraphics[width=26pc]{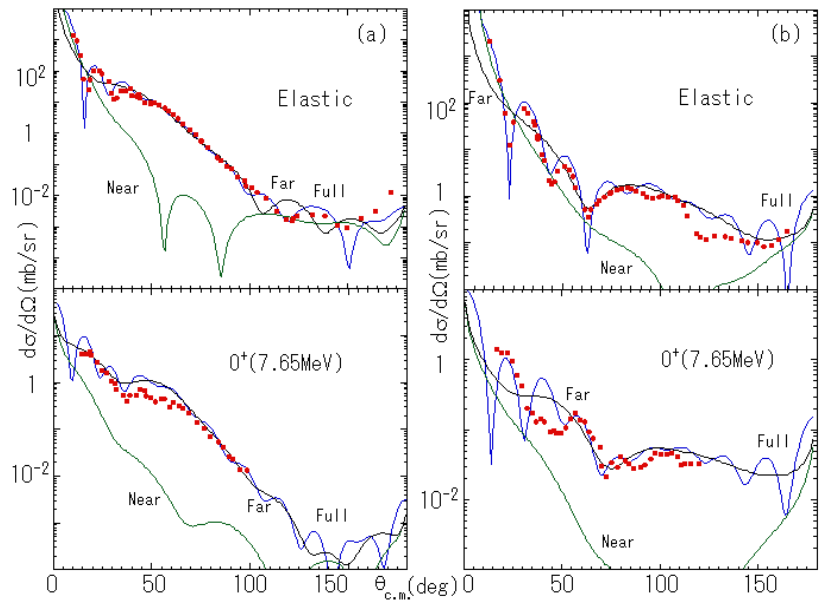}
\end{center}
\caption{Calculated cross sections  (solid lines), 
     farside (dotted lines) and nearside  (dashed lines)
contributions  in  $^3$He+$^{12}$C scattering at (a) 72 MeV and (b) 34.7 MeV
 are   compared  with the experimental data
 (points) \cite{Fujisawa1973,Demyanova1992}.}
\label{fig 2He3+C12:fig1}
\end{figure}

The calculated  scattering amplitude   is decomposed into  farside and nearside
 contributions following Fuller's prescription.
  At $E_L$=72 MeV the first Airy minimum $A1$ 
for elastic scattering  is not seen clearly in the experimental
 angular distribution. 
 On the other hand,
the $A1$ minimum for  the 0$^+_2$ state  is clearly seen in the farside cross sections
 because the minimum is shifted to a larger angle where  the nearside 
  contribution is  much smaller.
 At $E_L$= 34.7 MeV in  Fig.~2(b)  the Airy minimum $A1$
 appears at 60$^\circ$ for elastic scattering and 75$^\circ$ for the 0$^+_2$ state.
 The latter is  shifted to a much larger angle and the Airy minimum is not at all 
obscured 
 by the nearside contributions.
For the 0$^+_2$ state at $E_L$= 34.7 MeV  the nearside contributions 
are 
much smaller than the farside contributions compared with the elastic scattering
 case in the  wider range of angles.
Thus the difference of the refraction between the ground state 
and the 0$^+_2$ state is much
more clearly seen at 34.7 MeV than at 72 MeV.  

 The absorption
is incomplete for the 0$^+_2$ state and a more beautiful prerainbow Airy oscillation  is 
seen than for the elastic scattering.  The refractive effect of the 0$^+_2$
 state is more clearly seen  in the prerainbow  structure at the {\it low} incident
 energy
  region. Thus it is found
  that the incident 
$^3$He  is strongly refracted in the Hoyle state  in accordance with the picture
that the state has a large lens composed of  three alpha particles in a dilute
 density distribution.

 \section{Analysis of $\alpha$+$^{12}$C scattering}\label{analysis:sec3}
In a previous paper \cite{Ohkubo2004} we studied $\alpha$+$^{12}$C at the 
 high energy region above 139 MeV where a nuclear 
 rainbow with a typical falloff of the cross sections  appears in the 
angular distributions.  It was shown that for the Hoyle state the rainbow angle, 
and 
 therefore
 the associated Airy structure,  is shifted to a larger angle compared with 
 elastic scattering.

\begin{figure}[h]
\begin{center}
\includegraphics[width=26pc]{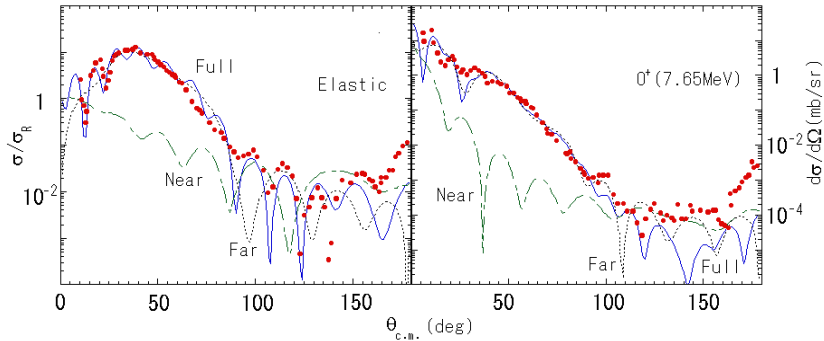}
\end{center}
\caption{ Calculated cross sections  (solid lines) 
 in $\alpha$+$^{12}$C  scattering
 at $E_L$=110 MeV are    decomposed into farside (dotted lines) and nearside 
 (dashed lines)
components and compared  with the experimental data
 (points) \cite{Demyanova2007}.}
\label{ang distri alpha+12C 110MeV:fig3}
\end{figure}

As shown in $^3$He+$^{12}$C scattering at 34.7 MeV, the  refractive effect can be 
seen more clearly at the lower energy region in the prerainbow oscillations.
Unfortunately for the $\alpha$+$^{12}$C system there are no experimental 
angular distributions for the Hoyle state in this low energy region. 
Also it has been  known that for the
$\alpha$+$^{12}$C system a global potential which reproduces  elastic scattering 
angular distributions in a wide range of incident energies including  the low
 energy region where ALAS (Anomalous Large Angle Scattering) appears has not been established.
Very recently Dem'yanova {\it et al.}  \cite{Demyanova2007} 
 measured   angular distributions  of inelastic scattering to the Hoyle state
 and elastic scattering   at 110 MeV. 
We have analyzed this data  in the same spirit and calculated results are shown in Fig.~3. 
The calculation reproduces characteristic features of the experimental data.
The rainbow angle for the Hoyle state is shifted to a larger angle compared
 with that for the ground state.
 The Airy minimum is clearly seen for the
 Hoyle state. 
These results confirm the findings of  the previous section in the   $^3$He+$^{12}$C scattering 
at 34.7 MeV.

As shown in Fig.~4, by  decomposing the total
 scattering cross sections at 110 MeV  into its partial wave contributions,
it is found that  
the peak of the partial cross sections for elastic scattering is located at orbital
 angular momentum $L=13$, while the peak for the inelastic scattering to the Hoyle
 state is located at $L=18$. This is also consistent with the picture that the 
refraction of inelastic scattering  in  the Hoyle state takes place at the larger 
radii than    the elastic scattering  \cite{Ohkubo2007B}. 
 
\begin{figure}
\begin{center}
  \includegraphics[width=24pc]{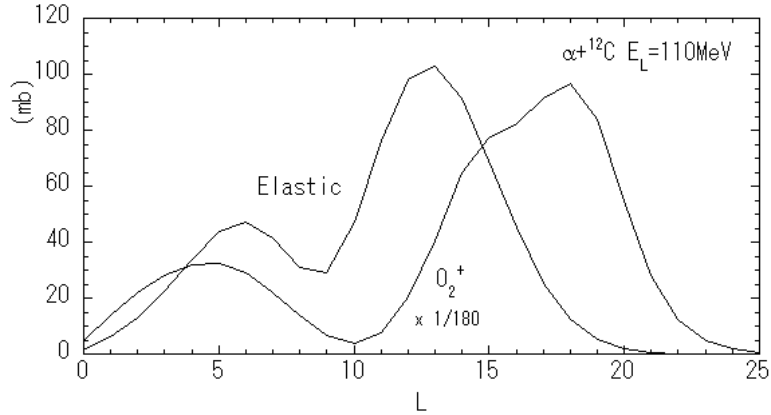}
\end{center}
  \caption{ Calculated partial cross sections for elastic $\alpha$+$^{12}$C 
scattering and
 inelastic scattering  in the    0$^+_2$ state at $E_L$=110 MeV are shown
 as a function of orbital  angular momentum.
}
\label{partial:fig4}
\end{figure}

 \section{Summary  }\label{discussion:sec4}
We have analyzed the angular distributions of prerainbow and rainbow scattering
 for the  $^3$He+$^{12}$C system  
 and rainbow scattering for the  $\alpha$+$^{12}$C  system in a  coupled channel method
 by using  a microscopic wave function for the Hoyle state that is almost equivalent to 
 the $\alpha$ particle condensate wave function. The experimental angular distributions
 are well reproduced by the calculations.
 It is found that refraction is much stronger for the inelastic scattering to the 
 Hoyle state than that for the elastic scattering and that the reaction takes place at 
the larger radii  for the Hoyle state than for  elastic scattering. These results are
 in agreement with the picture that the Hoyle 
 state has a large lens with  a dilute density distribution due to the $\alpha$
 particle condensation.

%\section*{Acknowledgments}
One of the authors (S.O.)   has been supported by a
 Grant-in-aid for Scientific Research
 of the Japan Society for Promotion of Science (No. 16540265).

\end{document}